\documentstyle[sprocl,psfig]{article}

\def\Journal#1#2#3#4{{#1} {\bf #2}, #3 (#4)}

\def\vp{{\bf p}}

\def\be{\begin{equation}}
\def\ee{\end{equation}}
\def\bea{\begin{eqnarray}}
\def\eea{\end{eqnarray}}

\def\begeq{\begin{equation}}
\def\endeq{\end{equation}}
\def\begeqar{\begin{eqnarray}}
\def\endeqar{\end{eqnarray}}
\def\mf{magnetic field}

\def\magn{mag\-ne\-tic}

\def\gm{gra\-vi\-to\-\magn}
\def\po{po\-lo\-idal}
\def\tor{to\-ro\-idal}

\def\1d#1{{1\over#1}}
\def\del{\partial}
\def\delt#1{{\del#1 \over \del t}}

\def\ddtau#1#2{{d#2 \over d\tau_{\rm #1}}}

\def\pri{^{\, '}}

\def\rme{^{\rm e}}

\def\o{\omega}
\def\ot{\tilde \omega}
\def\a{\alpha}
\def\ag{\alpha_{\rm g}}

\def\vbeta{\vec{\beta}}
\def\phid{\hat\phi}
\def\E{\vec{E}}
\def\Ep{\vec{E}_{\rm p}}

\def\B{\vec{B}}
\def\Bp{\vec{B}_{\rm p}}

\def\j{\vec{j}}
\def\jp{\vec{j}_{\rm p}}
\def\v{\vec{v}}
\def\vp{\vec{v}_{\rm p}}
\def\vph{v^{\phi}}
\def\vphid{v^{\phid}}
\def\S{\vec{S}}
\def\er{\vec{e}_r}

\def\e{\vec{e}}

\def\ephi{\vec{e}_{\phi}}
\def\ephid{\vec{e}_{\hat\phi}}
\def\nonu{\nonumber}
\def\n{\vec{\nabla}}
\def\A{\vec{A}}
\def\rot{\n \times}
\def\div{\n\cdot}
\def\tarrow#1{\buildrel\leftrightarrow\over #1}
\def\tens#1{\ifmmode\mathchoice{\mbox{$\sf\displaystyle#1$}}
{\mbox{$\sf\textstyle#1$}}
{\mbox{$\sf\scriptstyle#1$}}
{\mbox{$\sf\scriptscriptstyle#1$}}\else
\hbox{$\sf\textstyle#1$}\fi}
\newcommand{\subi}{_{\rm i}}
\newcommand{\sube}{_{\rm e}}

\newcommand{\me}{m_{\rm e}}

\newcommand{\rhoc}{\rho_{\rm c}}
\newcommand{\gammai}{\gamma_{\rm i}}
\newcommand{\gammae}{\gamma_{\rm e}}

\begin{document}

\title{GENERATION AND EVOLUTION OF MAGNETIC FIELDS IN THE GRAVITOMAGNETIC FIELD 
OF A KERR BLACK HOLE}

\author{RAMON KHANNA}

\address{Landessternwarte K\"onigstuhl,\\
D-69117 Heidelberg, Germany\\E-mail: Ramon.Khanna@lsw.uni-heidelberg.de}

\maketitle\abstracts{
I study the generation and evolution of magnetic fields in the plasma 
surrounding a rotating black hole. Attention is focused on effects of 
the gravitomagnetic potential. The gravitomagnetic force appears as 
battery term in the generalized Ohm's law. The generated magnetic field 
should be stronger than fields generated by the classical Biermann battery.
The coupling of the gravitomagnetic potential with 
electric fields appears as gravitomagnetic current in
Maxwell's equations. In the magnetohydrodynamic induction equation, this
current re-appears as source term for the poloidal magnetic field, which can 
produce closed magnetic structures around an accreting black hole.
In principle, even self-excited axisymmetric dynamo action is possible,
which means that Cowling's anti dynamo theorem does not hold in the Kerr 
metric.
Finally, the structure of a black hole driven current is studied.
}

\section{Introduction}
The influence of the gravitomagnetic field of a rotating compact object 
on electromagnetic fields has been studied for some 25 years (Wald 1974, 
Bicak \& Dvorak 1976). It was 
shown by Blandford \& Znajek (1977) that the coupling of the gravitomagnetic 
potential with a magnetic field results in an electromotive force. 
Currents driven by this 
electromotive force may extract rotational energy from a black hole.
Cast in the language of the 3+1 split of the Kerr metric, Maxwell's equations, 
together with the ingoing wave boundary condition for electromagnetic fields 
at the horizon, led to the 
{\it Membrane Paradigm} (Thorne et al. 1986), enforcing the analogy between 
a rotating black hole, immersed in an external magnetic field, with pulsars.
There are, however, some difficulties with this pulsar analogy, 
namely that (i) it is not easy to transport Poynting flux from close to the 
black hole's horizon out to infinity (Punsly \& Coroniti 1989, Punsly 1996), 
and (ii) a black hole does not carry its own magnetic field, not to mention 
kGauss fields. Magnetic fields have to be accreted onto the black hole 
either from the outer accretion disk, or must be generated in the plasma 
surrounding the black hole.

In this talk I address the question of the generation of magnetic fields 
by a battery operating in the plasma close to a rotating black hole 
and the evolution and the structure of magnetic fields that are brought in 
by an accretion disk. I show that the gravitomagnetic force may play a crucial 
role in the battery and that the coupling between the gravitomagnetic potential 
and an electric field is a source for the poloidal magnetic field, which 
may produce closed magnetic loops in the accreting plasma around the hole.
In principle, even an axisymmetric gravitomagnetic dynamo is possible, i.e. 
Cowling's anti-dynamo theorem is not valid in the Kerr-metric.

\section{The MHD description of an electron-ion plasma}
Here I summarize the relativistic definition of a plasma as center-of-mass 
fluid of its components, and the derivation of the generalized Ohm's law 
(Khanna 1998a).
The plasma is assumed to be a perfect fluid and is defined by the sum of the
ion and electron stress-energy tensors, which contain a collisional coupling 
term:
\begeq
        (\rho_{\rm m}\pri + p\pri)W^{\a}W^{\beta} + p\pri g^{\a\beta}
        \equiv T^{\a\beta} = \sum_{x=i,e}
	(\rho_{\rm mx}^{\rm x} + p_{\rm x}^{\rm x})
        W_{\rm x}^{\a}W_{\rm x}^{\beta} + p_{\rm x}^{\rm x} g^{\a\beta}
        + T^{\a\beta}_{\rm x\, coll}\; .
\label{defplas}
\endeq
Subscripts $i,e$ refer to ion and electron quantities, respectively. 
Superscripts denote the rest-frame in which the quantity is defined, 
where $\pri$ refers to the plasma rest-frame.
In the 3+1 split (into hypersurfaces of constant Boyer-Lindquist time $t$, 
filled with stationary zero angular momentum {\it fiducial observers})
\(T^{\a\beta} \) splits into the total density of mass-energy $\epsilon$ 
and momentum density $\S$
\begeq
        \epsilon \equiv (\rho_{\rm m}\pri + p\pri v^2)\gamma^2
        \approx \rho_{\rm m}\pri \gamma^2 
\qquad
        \S \equiv (\rho_{\rm m}\pri + p\pri)\gamma^2\v\approx \rho_{\rm m}\pri
        \gamma^2\v
\label{epsSdef}
\endeq
and the stress-energy tensor of 3-space with metric $\tarrow{h}$
\begeq
        \tarrow{T} \equiv (\rho_{\rm m}\pri +
        p\pri)\gamma^2 \v\otimes\v + p\pri\tarrow{h}
        \approx
        \rho_{\rm m}\pri\gamma^2 \v\otimes\v + p\pri\tarrow{h}\; .
\endeq
The approximate expressions hold for a `cold' plasma.
Adding the current density 4-vectors of each plasma component and
splitting into charge density and current density yields
\begeq
        \rho_{\rm c} \equiv \rho_{\rm ci} + \rho_{\rm ce} =
        Z e n_{\rm i}\gamma_{\rm i} - e n_{\rm e}\gamma_{\rm e}
\qquad
        \j \equiv \j_{\rm i} + \j_{\rm e} =
        Z e n_{\rm i}\gamma_{\rm i}\v_{\rm i}
        - e n_{\rm e}\gamma_{\rm e}\v_{\rm e}
    \; .
\endeq
All quantities resulting from the split are measured locally by FIDOs.
\subsection{The generalized Ohm's law in the 3+1 split of the Kerr metric}
In the `cold' plasma limit, the local laws of momentum conservation for 
each species can be re-written as equations of motion, which can then be 
combined to yield the generalized Ohm's law for an electron-ion plasma
\begeqar
        \frac{\j}{\sigma\gammae}&\approx&
        \E +\frac{Z n\subi\gammai}{n\sube\gammae}\v\times\B
        -\frac{\j\times\B}{e n\sube\gammae}
        +\frac{\n(\ag p\sube\rme)}{e n\sube\gammae\ag }
        +\frac{4\pi\gammae}{\o^2_{\rm pe}}\rhoc\pri\vec{g}
        +\frac{\rhoc\pri\gamma\v}{\sigma\gammae}
        \nonu\\
        &-&\frac{4\pi e(Z n\subi\gammae^2  - n\sube\gammai^2)}
               {\o_{\rm pe}^2 \gammae\gammai^2}\left(\ddtau{p}{(\gamma\v)}
               - \tarrow{H}\cdot(\gamma^2\v)\right) \; ,
\label{allgOhm}
\endeqar
with the conductivity \( \sigma = {e^2 n\sube}/{\me \nu_{\rm c}}\equiv 
\omega^2_{\rm pe}/4\pi\nu_{\rm c}\), the electron plasma frequency 
$\omega_{\rm pe}$, 
the factor of gravitational redshift $\ag$ (with \(\vec{g}=-\n\ln\ag\)) and 
the gravitomagnetic tensor field \(\tarrow{H}\equiv \ag^{-1} \n\vbeta\, .\) 
\(\vbeta = \beta^{\phi}\ephi\equiv -\o\ephi \) is the gravitomagnetic 
potential, which drags space into differential rotation with angular velocity 
$\o$. Note that, in the single fluid description, the 
gravitomagnetic force drives currents only, if the plasma is charged 
in its rest frame. $\tau_{\rm p}$ is the proper time in the plasma rest frame.
The derivation was made with the assumption that the species 
are coupled sufficiently strong that their bulk accelerations 
\begeq
        \ddtau{x}{(\gamma_{\rm x}\v_{\rm x})}\equiv
        \left[\frac{\gamma_{\rm x}}{\ag}\delt{}
        +\gamma_{\rm x}\left(\v_{\rm x}
        -\frac{\vbeta}{\ag}\right)\cdot\n\right](\gamma_{\rm x}\v_{\rm x})
\label{bulkacc}
\endeq
are synchronized. The same is required for the gravitomagnetic accelerations, 
i.e. 
\(|\tarrow{H}\cdot(\gammai^2\v\subi) -
        \tarrow{H}\cdot(\gammae^2\v\sube)|
         \ll
        |\tarrow{H}\cdot(\gammai^2\v\subi)|
\). 
If the MHD-assumption of ``synchronized accelerations''
is not made, Ohm's law contains further current acceleration terms, 
inertial terms and \gm\ terms (Khanna 1998a), which may be important for 
collisionless reconnection and particle acceleration along magnetic fields. 
This topic will be discussed elsewhere. 

In the limit of quasi-neutral plasma \((Z n\subi \approx  n\sube)\) and
\(\gammae\approx\gammai\approx\gamma\) Eq.~(\ref{allgOhm}) reduces to
\begeq
        \j \approx
        \sigma\gamma(\E +\v\times\B) -\frac{\sigma}{e n\sube}(\j\times\B)
        +\frac{\sigma}{e n\sube\ag}\n(\ag p\sube\rme) \; ,
\label{allgOhmqn}
\endeq
which contains all the terms, familiar from the non-relativistic generalized
Ohm's law, but no \gm\ terms.
\subsection{The gravitomagnetic battery}
The generation of magnetic fields by a plasma battery was originally devised 
by Biermann (1950) for stars. He showed that, if the centrifugal force 
acting on a rotating plasma does not possess a potential, the charge 
separation owing to the electron partial pressure cannot be balanced 
by an electrostatic field, and thus currents must flow and a magnetic field 
is generated. 

In Khanna (1998b) I have re-formulated Biermann's theory in 
the Kerr metric. The base of this battery theory is Ohm's law of 
eq.~(\ref{allgOhmqn}). Assuming that electrons and ions have
non-relativistic bulk velocities in the plasma rest frame, superscripts 
$i,e,\pri $ can be dropped.
With \(p = p\subi + p\sube = (n\subi + n\sube)kT\), the 
{\it impressed electric field} (IEF),
\(
        \E^{(i)} = {\n(\ag p\sube)}/{e n\gamma\ag}
\), 
can be re-expressed with the aid of the equation of motion for a 
`cold' quasi-neutral plasma to yield 
\begeq
        \E^{(i)} = \frac{m\subi}{(Z+1)e }
        \left(\gamma\vec{g} + \tarrow{H}\cdot(\gamma\v)
        -\ddtau{}{(\gamma\v)}\right)
        +\frac{Z \left(\j\times\B + (\v\cdot\j)\E\right) }
                {(Z+1) e n \gamma}\; .
\endeq
$\tau$ is the proper time in a FIDO frame; 
i.e. \(d / d\tau_{\rm p} = \gamma d / d\tau\).
The criterium for magnetic field generation is that 
$\rot{\ag\E^{(i)}}\ne 0\, .$ 
Here I restrict the discussion to 
the gravitomagnetic IEF $\E^{(i)}_{\rm gm}\; .$
The function part of \(\ag\E^{(i)}_{\rm gm}\) is
\begeqar
        \lefteqn{
        \left(\vbeta\cdot\n + \n\vbeta\, \cdot\right)(\gamma\v) =
        \left(\beta^i(\gamma v^j)_{|i} + \gamma\beta^{i|j} v_i\right) \e_j
        }\nonu\\
        &&= -\gamma\vph\ot^2\n\o
            -\o\left((\gamma v^r)_{,\phi}\er
                + (\gamma v^{\phi})_{,\phi}\ephi\right) \; ,
\label{Egm}
\endeqar
where $\ot = (h_{\phi\phi})^{1/2}$ and ${}_|$ denotes the covariant derivative 
in 3-space.
In axisymmetry \(\ag\E^{(i)}_{\rm gm}\) is clearly rotational,
unless some freak $\gamma$ should manage to make \(\gamma\vph\ot^2\) a function
of $\o$ alone. Thus the gravitomagnetic force drives a 
poloidal current and generates a toroidal magnetic field. 
Only if $\vph$ is non-axisymmetric, the \gm\ IEF drives \tor\
currents. The total IEF \((\ag\E^{(i)}_{\rm gm}+\ag\E^{(i)}_{\rm class})\) is 
likely to rotational in general. This will be clarified for specific 
velocity fields elsewhere.

In presence of a weak \po\ \mf\ the Biermann battery is limited due to 
modifications of the rotation law by the Lorentz force, rather than by 
ohmic dissipation. Then the contribution of the centrifugal 
force to the IEF becomes irrotational already at weak \tor\ fields 
(Mestel \& Roxburgh 1962). 
The gravitomagnetic battery term, on the other hand, 
is only linearly dependent on $\v$. The equilibrium field strength 
should therefore be higher than for the Biermann battery. 
\section{The MHD induction equation in the 3+1 split of the Kerr metric}
In this section I review the axisymmetric dynamo equations in the 3+1 split 
of the Kerr metric (Khanna \& Camenzind 1996a). 
Ohm's law is assumed to be of the standard form for a 
quasi-neutral plasma; Hall-term and IEF are neglected. 
Combining Maxwell's equations (Thorne et al. 1986) 
with Ohm's law yields the MHD induction equation 
\begeq
	\delt{\B} = \rot\left( (\ag\v\times\B ) - \frac{\eta}{\gamma}
                \left(\rot(\ag\B) + (\Ep\cdot\n\o)\ot\ephid\right) \right) 
		+ (\Bp\cdot\n\o)\ot\ephid\; . 
\label{MHDindeq}
\endeq
The term standing with the \magn\ diffusivity $\eta$ is the current density, 
which, via Amp\`ere's law, contains the coupling of the gravitomagnetic field 
with the electric field. In axisymmetry this is simply the shear of the 
\po\ electric field in the differential rotation of space, $\o$. Another 
induction term is the shear of the \po\ \mf\ by $\o$. This generates 
\tor\ \mf\ out of \po\ \mf\ even in a zero-angular-momentum flow.

\subsection{The gravitomagnetic dynamo}
Introducing the flux $\Psi$ of the \po\ \mf\ and the \po\ current $T$
\begeq
	\Psi = \1d{2\pi}\int\Bp\cdot d\A = \ot A^{\phid}
\qquad
	T = 2\int\ag\jp\cdot d\A = \ag\ot B^{\phid} \; ,
\endeq
where $A^{\phid}$ is the \tor\ component of the vector potential,
eq.~(\ref{MHDindeq}) splits into
\begeqar
       \delt{\Psi} &+& \ag (\vp\cdot\n)\Psi
        -\frac{\eta\ot}{\gamma} \vphid(\n\o\cdot\n\Psi)
        -\frac{\eta\ot^2}{\gamma}\div \left(\frac{\ag}{\ot^2}\n\Psi\right)
	\nonu\\ 
        &=&{} \frac{\eta\ot}{\gamma\ag}\left[ \left( T \vp -
              \frac{\eta}{\gamma}\n T\right)\times\ephid\right]\cdot\n\o
\label{dtPsi}
\endeqar
\begeqar
        \delt{T} &+& \ag (\vp\cdot\n)T
         +\ag\ot^2 T \left(\div\frac{\vp}{\ot^2} \right)
         - \ag\ot^2 \div\left(\frac{\eta}{\gamma\ot^2}\n T\right)
	\nonu\\ 
         &=&\ag\ot (\n\Psi\times\ephid) \cdot\n\Omega
        \; . \label{dtT}
\endeqar
These equations are the relativistic equivalent of the classical axisymmetric 
dynamo equations. It is important to note, however, that {\it no mean-field 
approach} was made, but $\Psi$ has source terms anyway. They result from 
$\Ep\cdot\n\o$. Obviously, Cowling's anti-dynamo theorem does not hold close to 
a rotating black hole. Growing modes of this gravitomagnetic dynamo were 
shown to exist for steep gradients of the plasma angular velocity $\Omega$ 
(N\`u\~nez 1997). For simple accretion scenarios, growing modes could not 
be found in kinematic numerical simulations (Khanna \& Camenzind 1996b). 
If, however, 
magnetic field is replenished by an outer boundary condition, the 
gravitomagnetic source terms generate closed loops around the black hole.
\subsection{The \mf\ structure in the accretion disk close to the hole}
\begin{figure}[]
\psfig{width=\textwidth,figure=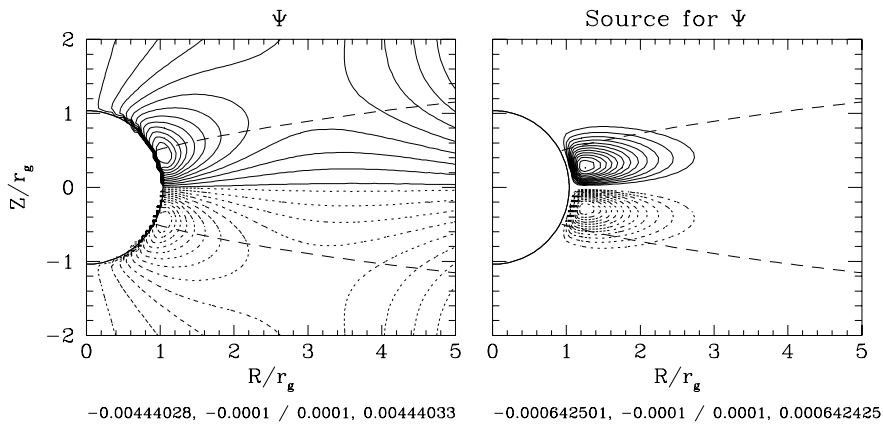,angle=-90}
\caption[ ]{Left: Contours of magnetic flux $\Psi$ showing a quadrupolar
magnetosphere of an accreting, rapidly rotating black hole ($a=0.998M$). 
Right: The
gravitomagnetic current as source of $\Psi$.
The disk is marked by long-dashed lines. Solid contours correspond to positive
values, short-dashed contours indicate negative values.
The range of contours is given below the boxes. }
\label{AG96}
\end{figure}

In the accretion disk, \mf\ may be advected into the near-horizon area, where 
gravitomagnetic effects may become important. This can be simulated by 
advection/diffusion boundary conditions for $F=\Psi ,\ T$
\begeq
	\frac{\del F}{\del n} + \frac{\gamma |v^{\hat r}|}{\eta}F = 
	\frac{\del F_{\rm out}}{\del n} 
	+ \frac{\gamma |v^{\hat r}|}{\eta}F_{\rm out}\; ,
\endeq
where $\del /\del n$ is the derivative along the outer boundary normal.
Figure~\ref{AG96} shows the stationary final state of a time-dependent 
simulation, in which $|B_{\rm p,out}| / |B_{\rm t,out}| = 1/50$. For such a 
dominantly \tor\ \mf\ the \gm\ source terms are strong enough to change the 
topology of $\Psi$. This may influence the efficiency of the electromagnetic 
extraction of rotational energy from the hole.
\subsection{Black Hole driven currents in accretion flows}
It was mentioned above that the shear of space does also induce a \tor\ \mf\ 
(cf. eq.~[\ref{MHDindeq}]). In eq.~(\ref{dtT}) this shear term is 
obscure, but still there, hidden in \((\n\Psi\times\ephid)\cdot\n\Omega\propto
\Bp\cdot\n \Omega = \Bp\cdot\n (\ag v^{\phi}+ \o)\). In a 
zero-angular-momentum flow $v^{\phi}=0$ (or, equivalently $\Omega = \o$) 
and thus the current $T$ is 
solely generated by the shear of space. Such a scenario can be used to 
assess, which fraction of the total current in a general accretion-ejection 
flow is driven by the hole. Figure~{\ref{AG98} shows the steady state of an 
initially homogenous vertical \mf\ that has been dragged onto the hole by 
radial accretion. The major part of the black hole driven current is 
confined within the ergosphere. The \tor\ field is of the same order as 
the \po\ field. The current's energy will be dissipated in the 
accretion flow and in the corona. 
In a realistic accretion-ejection flow, the plasma shear in the ergosphere 
of a rapidly rotating black hole will be similar to the shear of space. 
Thus the current system should have a structure similar to the current 
of Fig.~{\ref{AG98}. The energy extracted from the hole will therefore likely 
be deposited in the plasma close to the hole. 
\begin{figure}[]
\psfig{width=8cm,figure=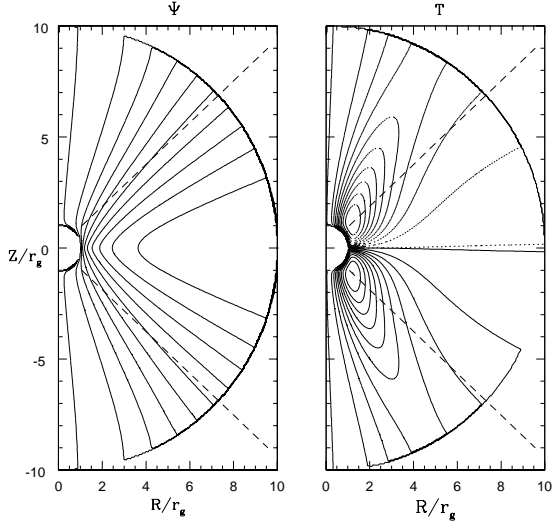,angle=-90}
\caption[ ]{
Poloidal magnetic field (left) and corresponding black hole generated current
(right) in a zero-angular-momentum accretion flow.
The specific angular memomentum of the black hole is $a=0.998 M$.
Dotted contours have negative values. Dashed
lines indicate the disk scale height $H$.}
\label{AG98}
\end{figure}
\section{Conclusions}
It was shown that a rotating black hole can generate magnetic fields in an 
initially un-magnetized plasma. In axisymmetry a plasma battery can only 
generate a \tor\ \mf , but then the coupling of the gravitomagnetic 
potential with \tor\ \mf s generates \po\ \mf s. 
Even an axisymmetric self-excited dynamo is theoretically possible, 
i.e. Cowling's theorem does not hold close to a Kerr black hole. 
Due to the joint action of \gm\ battery and \gm\ dynamo source term, 
a rotating black hole will always be surrounded by \po\ and \tor\ \mf s 
(probably of low field strength, though). The \gm\ dynamo source 
may generate closed \po\ \mf\ structures around the hole, which will 
influence the efficiency of the Blandford-Znajek mechanism. 

The ``shear-of-space'' driven fraction of a global current can be assessed by 
a kinematic simulation of a zero-angular-momentum flow. The major part of the 
resulting current system is generated and closed in the corona near the hole. 
In a realistic accretion-ejection flow, the plasma shear in the ergosphere
of a rapidly rotating black hole will be similar to the shear of space.
The current system should therefore have a similar structure as in the 
example shown here, which means that the energy extracted from 
the hole is likely to be deposited in the disk corona.

\section*{Acknowledgments}
This work was supported by the Deutsche Forschungsgemeinschaft (SFB 328).

\end{document}